\documentclass{article}
\usepackage[dvips]{graphicx}  

\begin{document}

\newcommand{\rum}{\rule{0.5pt}{0pt}}

\newcommand{\rub}{\rule{1pt}{0pt}}

\newcommand{\rim}{\rule{0.3pt}{0pt}}

\newcommand{\numtimes}{\mbox{\raisebox{1.5pt}{${\scriptscriptstyle \times}$}}}

\renewcommand{\refname}{References}


\begin{center}

{\Large\bf Optical-Fiber Gravitational Wave Detector: Dynamical 3-Space Turbulence Detected \rule{0pt}{13pt}}\par

\bigskip

Reginald T. Cahill \\ 

{\small\it School of Chemistry, Physics and Earth Sciences, Flinders University,

Adelaide 5001, Australia\rule{0pt}{13pt}}\\

\raisebox{-1pt}{\footnotesize E-mail: Reg.Cahill@flinders.edu.au}\par

\bigskip\smallskip

{\small\parbox{11cm}{%

Preliminary results from an optical-fiber gravitational wave interferometric detector are reported.  The detector is very small, cheap and simple to build and operate. It is assembled  from readily available opto-electronic components. A parts list  is given. The detector can operate in two modes: one in which only instrument noise  is detected, and data from a 24 hour period is reported for this mode, and in a 2nd  mode in which the gravitational waves are detected as well, and data from a 24 hour period is analysed.  Comparison shows that the instrument has a high S/N ratio.  The frequency spectrum of the gravitational waves shows  a pink noise spectrum, from  0 to 0.1Hz.
\rule[0pt]{0pt}{0pt}}}\bigskip

\end{center}

\section{Introduction}
\begin{figure}
\vspace{2mm}
\hspace{20mm}\includegraphics[scale=0.92]{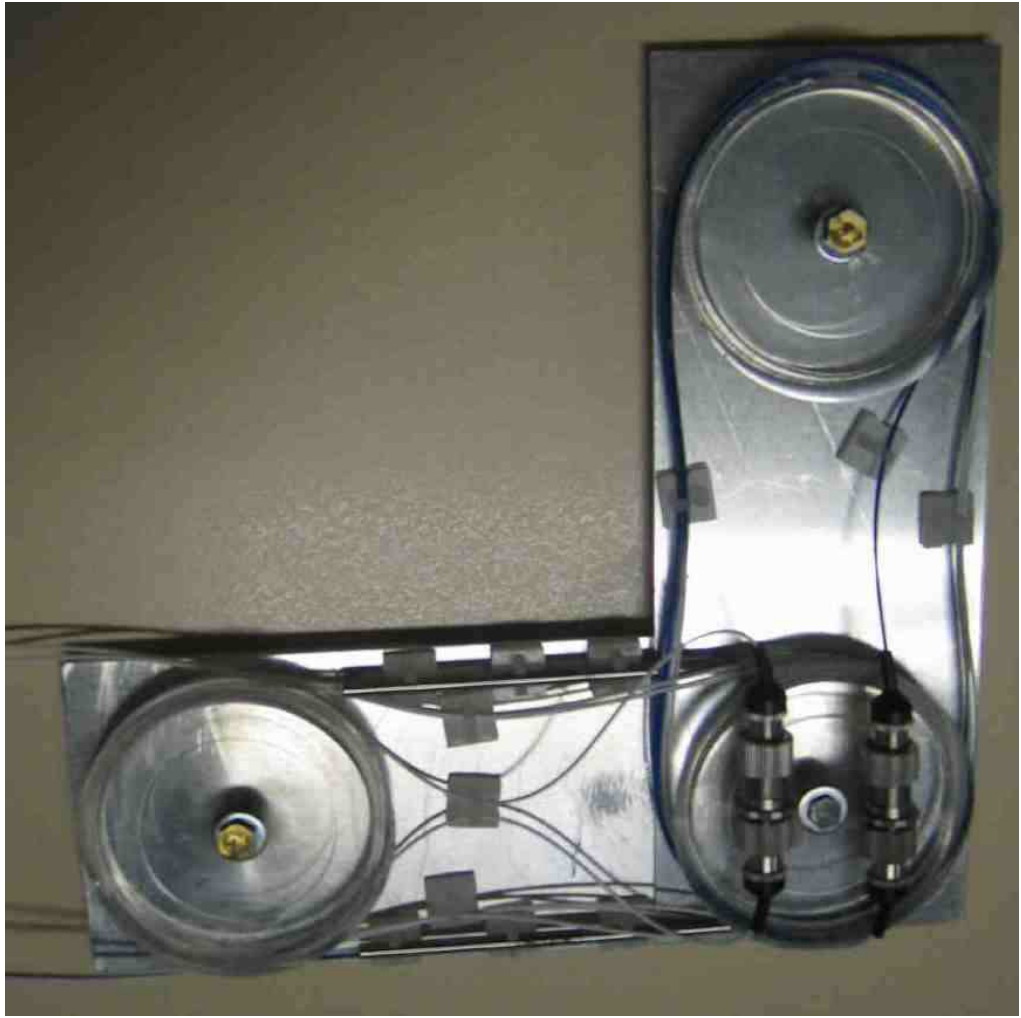}
\vspace{0mm}
\caption{\small{Photograph of the detector showing the fibers forming the two orthogonal arms. See Fig. \ref{fig:schematic} for the schematic layout.  The beam splitter and joiner are the two  small stainless steel cylindrical tubes. The two FC to FC mating sleeves   are physically adjacent, and the fibers can be re-connected to change from Mode A (Active detector - gravitational wave and device noise detection)  to Mode B (Background -  device noise measurements only).   The overall dimensions are 160mm$\times$160mm. The 2$\times$2 splitter and joiner each have two input and two output fibers, with one not used.  In operation the detector must be isolated from temperature fluctuations, particularly those that could be caused by air convection currents.
  \label{fig:photo}}}
\end{figure}

Preliminary results from an optical-fiber gravitational wave interferometric detector are reported.  The detector is very small, cheap and simple to build and operate, and is shown in Fig.\ref{fig:photo}.  It is assembled  from readily available opto-electronic components, and is suitable for amateur and physics undergraduate laboratories. A parts list  is given. The detector can operate in two modes: one in which only instrumental noise  is detected, and the 2nd in which the gravitational waves are detected as well.  Comparison shows that the instrument has a high S/N ratio.  The frequency spectrum of the gravitational waves shows  a pink noise spectrum, from 0 to 0.1Hz.  The interferometer is 2nd order in v/c and is analogous to a Michelson interferometer. Michelson interferometers in vacuum mode cannot detect the light-speed anisotropy effect or the gravitational waves manifesting as  light-speed anisotropy fluctuations.  The design and operation as well as preliminary data analysis are reported here so that duplicate detectors may be constructed to study   correlations over various  distances.  The source of the gravitational waves is unknown, but a 3D multi-interferometer detector will  soon be able to  detect directional characteristics of the waves.

\begin{figure*}[t]
\vspace{30mm}
\hspace{7mm}
\setlength{\unitlength}{1.5mm}
\hspace{22mm}\begin{picture}(0,0)
\thicklines
\put(4,0.0){\line(1,0){10}}
\put(4,2){\line(1,0){10}}
\put(4,0){\line(0,1){2}}
\put(14,0){\line(0,1){2}}
\put(6,5.5){\bf laser-}
\put(6,+3){\bf diode}

\put(18,1.25){\line(1,0){10}}
\put(18,1.15){\line(1,0){10}}
\put(18,1.05){\line(1,0){10}}
\put(18,0.95){\line(1,0){10}}
\put(18,0.85){\line(1,0){10}}
\put(18,0.75){\line(1,0){10}}
\put(18,0.75){\line(0,1){0.5}}
\put(28,0.75){\line(0,1){0.5}}
\put(19,6.5){\bf beam-}
\put(19,4.0){\bf splitter}
\put(14,1.0){\line(1,0){4}}
\put(14,1.0){\vector(1,0){3}}

\put(4,-5){\line(1,0){4}}
\put(4,-3){\line(1,0){4}}
\put(4,-5){\line(0,1){2}}
\put(8,-5){\line(0,1){2}}
\put(4,-8){\bf photo-}
\put(4,-10.5){\bf diode}

\put(-4,-5){\line(1,0){4}}
\put(-4,-3){\line(1,0){4}}
\put(-4,-5){\line(0,1){2}}
\put(0,-5){\line(0,1){2}}
\put(-4,-8){\bf DSO}
\put(0,-4){\line(1,0){4}}

\put(18,-3.75){\line(1,0){10}}
\put(18,-3.85){\line(1,0){10}}
\put(18,-3.95){\line(1,0){10}}
\put(18,-4.05){\line(1,0){10}}
\put(18,-4.15){\line(1,0){10}}
\put(18,-4.25){\line(1,0){10}}

\put(18,-4.25){\line(0,1){0.5}}
\put(28,-4.25){\line(0,1){0.5}}
\put(19,-7.0){\bf beam-}
\put(19,-9.5){\bf joiner}
\put(8,-4){\line(1,0){10}}
\put(16,-4){\vector(-1,0){3}}
\put(28,4.4){\oval(40,6)[r,b]}
\put(50.5,4.2){\oval(5,40)[l,t]}
\put(50,-1){\oval(5,50.4)[t,r]}
\put(50.3,-0.9){\oval(4.5,4.5)[r,b]}
\put(50.5,-0.6){\oval(4.,5.)[l,b]}
\put(50.5,-0.6){\oval(4.,48)[l,t]}
\put(50.5,-2.6){\oval(3.,52)[r,t]}
\put(28,-1.65){\oval(48,4)[b,r]}

\put(28,1.4){\vector(1,0){5}}
\put(28,0.7){\vector(1,0){7}}
\put(28,-2.0){\oval(90,5.5)[r,t]}
\put(49.0,-1.85){\oval(48.0,3)[b,r]}
\put(50.0,-1.40){\oval(5,3.9)[l,b]}
\put(70.0,-1.7){\oval(45,4)[l,t]}
\put(69.5,-2.2){\oval(8.0,5)[t,r]}
\put(28,-1.8){\oval(91,5)[b,r]}

\put(38,-3.7){\vector(-1,0){5}}
\put(38,-4.35){\vector(-1,0){7}}
\put(48.3,-0.3){\line(1,0){0.5}}
\put(48.3,-1.3){\line(1,0){0.5}}
\put(48.3,-2.3){\line(1,0){0.5}}
\put(48.3,-2.3){\line(0,1){2}}
\put(48.8,-2.3){\line(0,1){2}}
\put(36,-7){\bf FC mating sleeves}
\put(38,10){\bf ARM 1}
\put(47.3,-0.3){\line(1,0){0.5}}
\put(47.3,-1.3){\line(1,0){0.5}}
\put(47.3,-2.3){\line(1,0){0.5}}
\put(47.3,-2.3){\line(0,1){2}}
\put(47.8,-2.3){\line(0,1){2}}
\put(58,2.0){\bf ARM 2}

\put(52.0,-9.0){\vector(1,0){19}}
\put(58.0,-9.0){\vector(-1,0){8}}
\put(58,-8.0){\bf 100mm}

\put(70,-2.0){\bf o}
\put(49.5,21.0){\bf o}

\put(49.2,-1.0){\bf x}
\put(49.2,-2.6){\bf x}
\put(45.2,-1.0){\bf w}
\put(45.2,-2.6){\bf w}

\end{picture}
\vspace{15mm}
\caption{  \small{Schematic layout of the interferometric optical-fiber  light-speed anisotropy/gravitational wave detector (in Mode {\bf A}).  Actual detector   is shown  in Fig.\ref{fig:photo}, with ARM2 located to the left, so as to reduce lengths of fiber feeds and overall size.  Coherent 650nm light from the laser diode is split into two 1m length single-mode  polarisation preserving fibers by the beam splitter. The two fibers take different directions, ARM1 and ARM2, after which the light is recombined in the beam joiner, which also has 1m length fibers, in which the phase differences lead to interference effects  that are indicated by the outgoing light intensity, which is measured in the photodiode, and then recorded in the Digital Storage Oscilloscope (DSO).  In Mode {\bf A} the optical fibers are joined $x-x$ and $w-w$ at the  FC to FC mating sleeves, as shown.  In the actual layout the fibers make four loops in each arm, and the length of one straight section is 100mm, which is the center to center spacing of the plastic turners, having diameter = 52mm, see Fig.\ref{fig:photo}.  The two FC to FC mating sleeves are physically adjacent and  by re-connecting the fibers as $x-w$ and $w-x$ the light paths can be made symmetrical wrt the arms, giving Mode {\bf B}, which only responds to device noise - the {\bf B}ackground mode. In Mode {\bf A} the detector is {\bf A}ctive, and responds to both flowing 3-space and device noise. The relative travel times, and hence the output light intensity, are affected by the fluctuating speed and direction of the flowing 3-space, by affecting differentially the speed of the light, and hence the net phase difference between the two arms.   }}
 \label{fig:schematic}
\end{figure*}
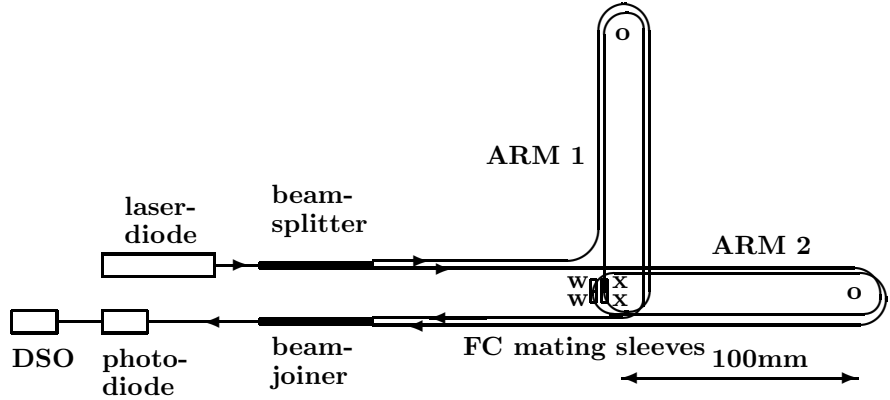

\section{Light Speed Anisotropy}
In 2002 it was reported \cite{MMCK,MMC} that   light-speed anisotropy had been detected repeatedly since the Michelson-Morley experiment of 1887 \cite{MM}.  Contrary to popular  orthodoxy they reported a light-speed anisotropy up to 8km/s based on their analysis  of their observed fringe shifts. The Michelson-Morley experiment was everything except {\it null}.  The deduced speed was based on Michelson's Newtonian-physics calibration for the interferometer. In 2002 the necessary special relativity effects and the effects of the air present in the light paths were first taken into account in calibrating the interferometer. This reanalysis showed that the actual observed fringe shifts corresponded to a  very large  light-speed anisotropy, being in excess of 1 part in 1000 of c = 300,000km/s. The existence of this light-speed anisotropy is not  in conflict with the successes of Special Relativity, although it is in conflict with Einstein's postulate that the speed of light is invariant.   This large light-speed anisotropy had gone unnoticed throughout the twentieth century, although we now know that it was detected in seven experiments, ranging from five  2nd order in v/c  gas-mode Michelson-interferometer experiments \cite{MM,Miller,C5,C6,C7} to two 1st order in v/c one-way RF  coaxial cable travel-speed measurements   using atomic clocks   \cite{Torr,DeWitte}.   In 2006 another RF travel time coaxial cable experiment was performed \cite{Coax}. All  eight light-speed anisotropy experiments agree \cite{Book,Review}. Remarkably  five of these experiments \cite{MM,Miller,Torr,DeWitte,Coax} reveal pronounced {\it gravitational wave} effects, where the meaning of this term is explained below. In particular detailed analysis of the Michelson-Morley fringe shift data shows that they not only detected a large light-speed anisotropy, but that their data also reveals large wave effects \cite{Review}.  The reason why their interferometer could detect these phenomena was that the light paths passed through air; if a Michelson interferometer is operated in vacuum then changes in the geometric light-path lengths exactly cancel the Fitzgerald-Lorentz arm-length contraction effects.  This cancellation is incomplete when a gas is present in the light paths.   So modern vacuum Michelson interferometers are incapable of detecting the large light-speed anisotropy or the large gravitational waves.  Here we detail the construction  of a simple optical-fiber light-speed anisotropy detector, with the main aim being to record and characterise the gravitational waves. These waves reveal a fundamental aspect to reality that is absent in the prevailing models of reality.

\section{Dynamical 3-Space and Gravitational Waves}
The light-speed anisotropy experiments reveal that a dynamical 3-space exists, with the speed of light being $c$ only wrt to this space: observers in motion `through' this 3-space detect that the speed of light is in general different from $c$, and is different in different directions.  The notion of a dynamical 3-space is reviewed in \cite{Book,Review}.  The dynamical equations for this 3-space are now known and involve a velocity field ${\bf v}({\bf r},t)$, but where only relative velocities are observable.  The coordinates ${\bf r}$ are relative to a non-physical mathematical embedding space. These dynamical equations involve Newton's gravitational constant $G$ and the fine structure constant $\alpha$.  The discovery of this dynamical 3-space then required a generalisation of the Maxwell, Schr\"{o}dinger and Dirac equations. In particular these equations showed that the phenomenon of gravity is a wave refraction effect, for both EM waves and quantum matter waves \cite{Review,Schrod}. This new physics has been confirmed by explaining the origin of gravity, including the Equivalence Principle, gravitational light bending and lensing, bore hole $g$ anomalies, spiral galaxy rotation anomalies (so doing away with the need for dark matter), black hole mass systematics, and also giving an excellent parameter-free fit to the supernovae and gamma-ray burst Hubble expansion data \cite{Hubble} (so doing away with the need for dark energy). It also predicts a novel spin precession effect in the GPB satellite gyroscope experiment \cite{GPB}. This physics gives an explanation for the successes of the Special Relativity formalism, and the geodesic formalism of General Relativity.  The wave effects already detected correspond to fluctuations in the 3-space velocity field ${\bf v}({\bf r},t)$, so they are really 3-space turbulence or wave effects.  However they are better known, if somewhat inappropriately as `gravitational waves' or `ripples' in `spacetime'.  Because the 3-space dynamics gives a deeper understanding of the spacetime formalism, we now know that the metric of the induced spacetime, merely a mathematical construct having no ontological significance, is related to ${\bf v}({\bf r},t)$ according to \cite{Book,Review}
\begin{equation}
ds^2=dt^2 -(d{\bf r}-{\bf v}({\bf r},t)dt)^2/c^2
=g_{\mu\nu}dx^{\mu}dx^\nu
\label{eqn:Eqn1}\end{equation}
The gravitational acceleration of matter, and of the structural patterns  characterising  the 3-space, is given by \cite{Review,Schrod}
\begin{equation}
{\bf g}=\frac{\partial {\bf v}}{\partial t}+({\bf v}.\nabla ){\bf v}
\label{eqn:acceln}
\end{equation}
and so fluctuations in  ${\bf v}({\bf r},t)$ may or may not manifest as a gravitational force.
The general characteristics of   ${\bf v}({\bf r},t)$ are now known following the detailed analysis  of the eight experiments noted above, namely its average speed, over an hour or so, of some 420$ \pm $30km/s, and direction RA= $5.5 \pm 2^{hr}$, Dec=$70 \pm 10^o$S, together with large wave/turbulence effects.  The magnitude of this turbulence depends on the timing resolution of each particular experiment, and here we report  that the speed fluctuations are very large, as also seen in \cite{Coax}.  Here we employ a new  detector design that enables a detailed  study of  ${\bf v}({\bf r},t)$, and with small timing resolutions.
A key experimental test of the various detections of ${\bf v}({\bf r},t)$ is that the data  shows that the time-averaged  ${\bf v}({\bf r},t)$ has a direction that  has a specific Right Ascension and Declination as given above, i.e. the time for say a maximum averaged speed depends on the local sidereal time, and so varies considerably throughout the year, as do the directions to  all astronomical  processes/objects. This sidereal effect constitutes an absolute proof that the direction of  ${\bf v}({\bf r},t)$ and the accompanying wave effects are real `astronomical' phenomena, as there is no known earth-based effect that can emulate the sidereal effect.  

\begin{figure*}
\vspace{-4mm}
\hspace{-4mm}\includegraphics[scale=0.85]{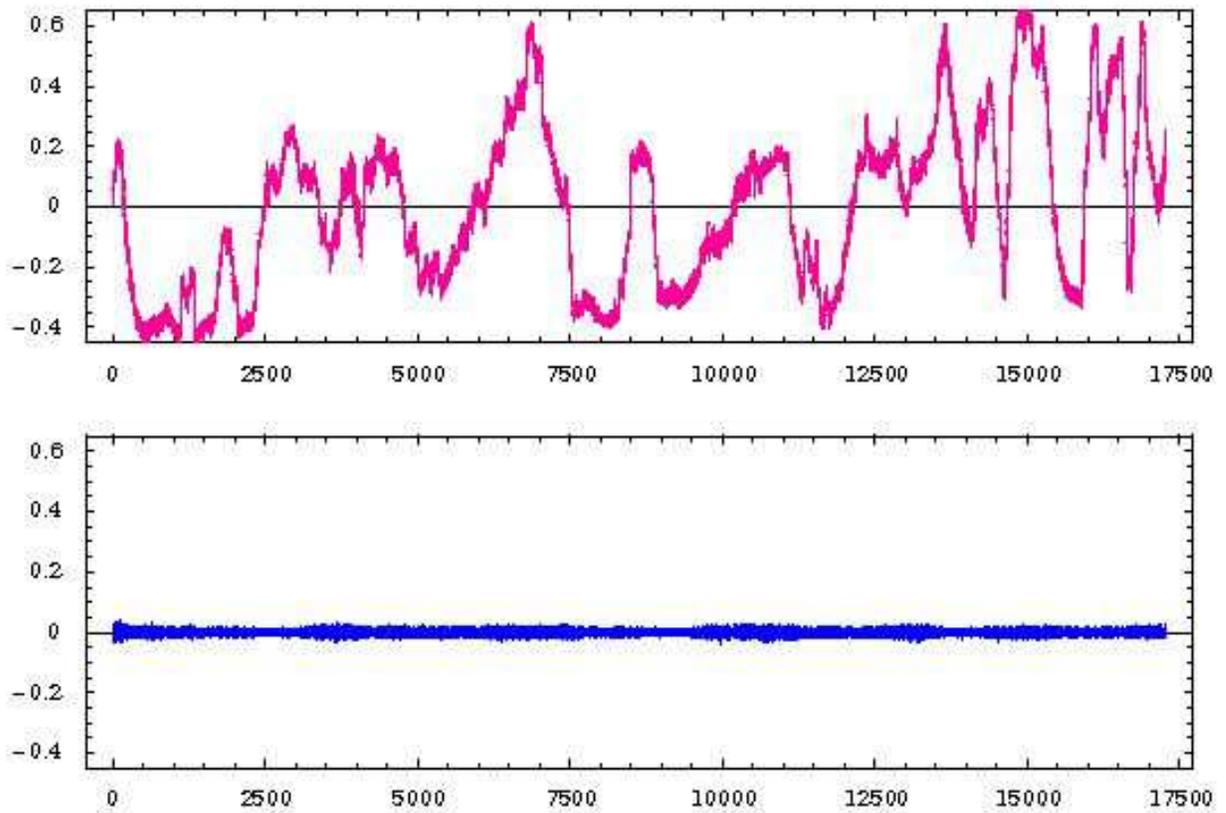}
\vspace{-9mm}
\caption{\small{Photodiode voltages over a 24 hour period with data recording every 5s, with the detector arms orientated in a NS-EW direction and horizontal.  Upper plot  (red) is for detector in Mode A, i.e responding to 3-space dynamics and instrument noise, while lower plot (blue)  is for Mode B in which detector only responds to instrumental noise, and demonstrates the high S/N ratio of the detector.  The lower plot is dominated by higher frequency noise, as seen in the frequency spectrum in  Fig. \ref{fig:FT}.     A selection of the above data over a 1 hour time interval, from time steps 4900 to 5620, is shown in Fig.  \ref{fig:ShortTime} indicating details of the 3-space waves. \label{fig:AllData}}}
\end{figure*}

\section{Gravitational Wave Detector}
To measure  ${\bf v}({\bf r},t)$ has been difficult until now. The early experiments used gas-mode Michelson interferometers, which involved the visual observation of small fringe shifts as the relatively large devices were rotated.  The RF coaxial cable experiments had the advantage of permitting electronic recording of the RF travel times, over 500m \cite{Torr} and 1.5km \cite{DeWitte}, by means of two or more atomic clocks, although the experiment reported in \cite{Coax} used a novel technique that enable the coaxial cable length to be reduced to laboratory size.  The new detector design herein has the  advantage of electronic recording as well as high precision because the travel time differences in the two orthogonal fibers employ light interference effects, with the interference effects taking place in an optical beam-joiner, and so no optical projection problems arise.  The device is very  small, very cheap and easily assembled from readily available opto-electronic components. The schematic  layout of the detector is given in Fig. \ref{fig:schematic}, with a detailed description in the figure caption.  The detector relies on the phenomenon where the 3-space velocity ${\bf v}({\bf r},t)$ affects differently the light travel times in the  optical fibers, depending on the projection of ${\bf v}({\bf r},t)$ along the fiber directions. The differences in the light travel times are measured by means of the interference effects in the beam joiner.  However at present the calibration constant $k$ of the device is not yet known, so it is not yet known what speed corresponds to the measured time difference $\Delta t$, although comparison with the earlier experiments gives a  guide.  In general we expect 
 \begin{equation}
\Delta t=k^2\frac{Lv_P^2}{c^3}\cos\bigl(2(\theta-\psi)\bigr)
\label{eqn:times}\end{equation}   
where $k$ is the instrument calibration constant. For gas-mode Michelson interferometers $k$ is known to be given by $k^2\approx n^2-1$, where $n$ is the refractive index of the gas. Here $L=4\times100$ mm is the effective arm length,  achieved by having four loops of the fibers in each arm, and $v_P$ is the projection of ${\bf v}({\bf r},t)$ onto the plane of the detector. The angle $\theta$ is that of the arm relative to the local meridian, while $\psi$ is the angle of the projected velocity, also relative to the local meridian. 
A photograph of the prototype detector is shown in Fig. \ref{fig:photo}.

\begin{table}
{\vspace{3mm}
\hspace{35mm}\begin{tabular}{| l|c |c|c|c|c|} 
\hline\hline
 {\bf Parts}  & {\bf Thorlabs  } \\ 
 &http://www.thorlabs.com/ \\ \hline 
1x Si Photodiode Detector/  & PDA36A or PDA36A-EC \\ 
Amplifier/Power Supply &  select for local AC voltage \\ \hline
1xFiber Adaptor for above & SM1FC \\ \hline
1xFC Fiber Collimation Pkg & F230FC-B  \\  \hline
1xLens Mounting Adaptor & AD1109F  \\ \hline
 2xFC to FC Mating Sleeves  &ADAFC1 \\ \hline
  2x  2x2 Beam Splitters  &  FC632-50B-FC \\ \hline
  Fiber Supports & PFS02 \\  \hline\hline
  & {\bf Midwest Laser Products }  \\ 
   &http://www.midwest-laser.com/\\ \hline
 650nm Laser  Diode Module & VM65003\\ \hline\hline
    LDM Power Supply/3VDC  & {\bf Local Supplier or Batteries} \\ \hline \hline
  BNC 50$\Omega$ coaxial cable    & {\bf Local Supplier } \\ \hline \hline
  & {\bf PoLabs}   \\ \hline
 PoScope USB DSO  &  http://www.poscope.com/\\ 
 \hline\hline
\end{tabular}}
\label{tab:table}
\vspace{-3mm}\caption{\small{List of parts and possible suppliers  for the detector. The FC Collimation Package and Lens Mounting Adaptor together permit the coupling of the Laser Diode Module to the optical fiber connector.  This requires unscrewing the lens from the Laser Diode Module and screwing the diode into above and making judicious adjustment to maximise light coupling. The  coaxial cable is required to connect the photodiode output to the DSO. Availability of a PC to host the USB DSO is assumed. The complete detector will cost $\approx\$1100$USD.}}
\end{table}

A key component is the light source, which can be the laser diode listed in the Table of parts. This has a particularly long coherence length, unlike most cheap laser diodes, although the data reported herein used a more expensive He-Ne laser.  The other key components are the  fiber beam splitter/joiner, which split the light into the fibers for each arm, and recombine the light for phase difference measurements by means of the fiber-joiner and photodiode detector and amplifier.  A key feature of this design is that the detector can operate in two different modes.  In Mode {\bf A} the detector is {\bf A}ctive, and responds to both flowing 3-space and device noise.  Because the two fiber coupler (FC) mating sleeves are physically adjacent a re-connection of  the fibers  at the two mating sleeves  makes the light paths  symmetrical wrt the arms, and then the detector only responds to device noise; this is the  {\bf B}ackground mode.   The data stream may be mostly cheaply  recorded   by a PoScope USB Digital Storage Oscilloscope (DSO) that runs on a PC.  

The interferometer operates by detecting the travel time difference between the two arms  as given by (\ref{eqn:times}). The cycle-averaged light intensity emerging from the beam joiner is given by
\begin{eqnarray}
I(t)&\propto&|{\bf E}_1e^{i\omega t}+{\bf E}_2e^{i\omega( t+\tau +\Delta t})|^2 \nonumber \\
&=&|{\bf E}|^2\cos\left(\frac{\omega(\tau+\Delta t)}{2}\right)^2   \nonumber \\
&\approx& a+b\Delta t
\label{eqn:intensity}\end{eqnarray}
Here ${\bf E}_i$ are the electric field amplitudes  and have the same value as the fiber splitter/joiner are 50\%-50\% types,  and having the same direction because polarisation preserving fibers are used,   $\omega$ is the light angular frequency and $\tau$ is a travel time difference caused by the light travel times not being identical, even when $\Delta t=0$, mainly because the various splitter/joiner fibers will not be identical in length. The last expression follows because $\Delta t$ is small, and so the detector operates in a linear regime, in general, unless $\tau$ has  a value equal to modulo($T$), where $T$ is the light period.  The main temperature effect in the detector is that $\tau$ will be temperature dependent. The photodiode detector output voltage  $V(t)$  is proportional to $I(t)$, and so finally linearly related to $\Delta t$.  The detector calibration constants $a$ and $b$ depend on $k$ and $\tau$,  and are unknown at present, and indeed $\tau$ will be instrument dependent. The results reported herein show that the value of the calibration constant $b$ is not given by using the effective refractive index  of the optical fiber in (\ref{eqn:times}), with $b$ being much smaller than that calculation would suggest. This is in fact very fortunate as otherwise the data would be affected by the need to use the cosine form in (\ref{eqn:intensity}), and thus would suffer from modulo effects.  It is possible to determine the voltages for which  (\ref{eqn:intensity}) is in the non-linear regime by spot heating a segment of one fiber by touching with a finger, as this produces many full fringe shifts.

By having three mutually orthogonal optical-fiber interferometers it is possible to deduce the vectorial direction of ${\bf v}({\bf r},t)$, and so determine, in particular,  if the pulses have any particular direction, and so a particular source. The simplicity of this device means that an international network of detectors may be easily set up, primarily to test for correlations in the waveforms.

\section{Data Analysis}
\begin{figure}
\vspace{0mm}
\hspace{15mm}\includegraphics[scale=0.62]{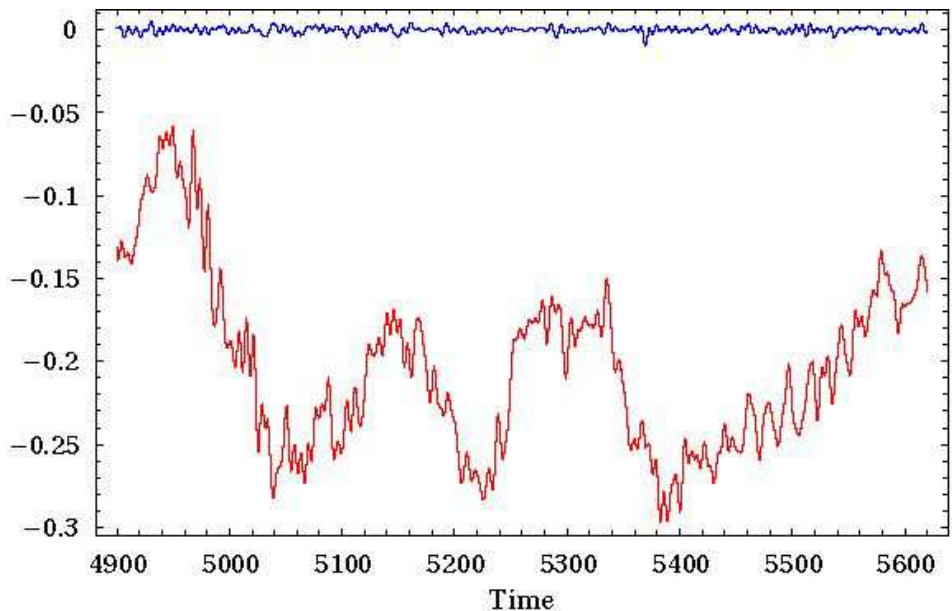}
\vspace{-3mm}
\caption{\small{Lower  plot (red) shows the time series data over a 1 hour period, from time steps 4900 to 5620 in Fig. \ref{fig:AllData}, showing the wave forms present in Fig. \ref{fig:AllData} in greater detail.  Similar complex wave forms were seen in \cite{Coax}. These plots were re-constructed from the FT after band passing the frequencies $(1 - 3000) \times1.16\times10^{-5}\mbox{Hz}=(0.000116-0.034)$Hz to reduce the instrument noise component, which is very small  as shown in upper plot (blue).  \label{fig:ShortTime}}}
\end{figure}

Photodiode voltage readings from the detector in Mode {\bf A} on July 11, 2007, from approximately 12:30pm local time for 24 hours, and in Mode {\bf B} June 24 from 4pm local time for 24 hours, are shown in Fig. \ref{fig:AllData}, with the zero frequency component removed. The photodiode output voltages were  recorded every 5s.  Most importantly the data are very different, showing that only in Mode {\bf A} are gravitational waves detected, and with a high S/N ratio. A 1 hour time segment of that data is  shown in Fig. \ref{fig:ShortTime}. In that plot the higher frequencies have been filtered out from both data time series, showing the exceptional S/N ratio that can be achieved.

\begin{figure}
\hspace{15mm}\includegraphics[scale=0.63]{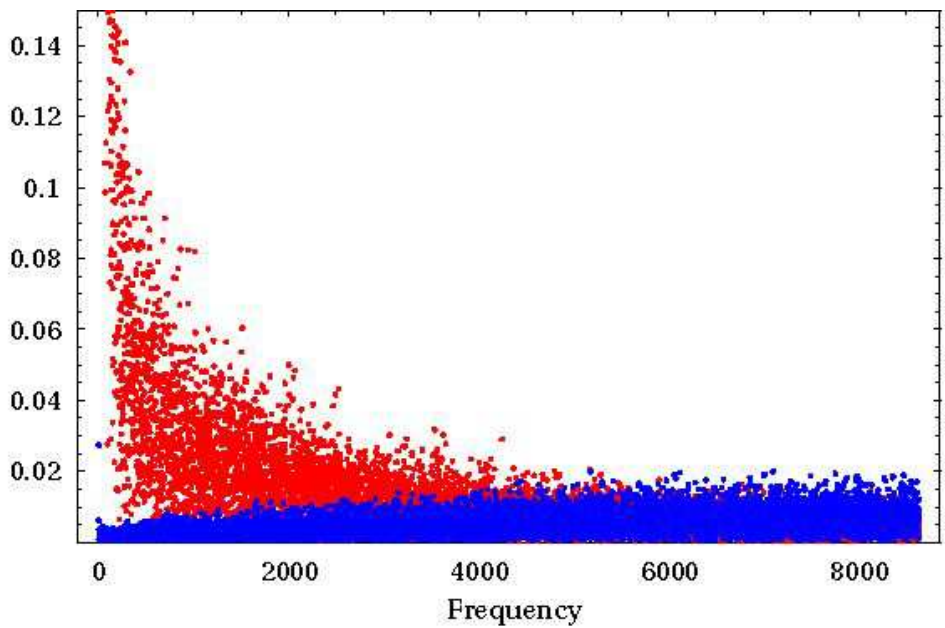}
\vspace{-3mm}
\caption{\small{ Two plots of $|\tilde{V}_s|$  from  Fast Fourier Transforms of the photodiode detector voltage $V_r$ at 5 second intervals  for 24 fours.    Frequency step corresponds to $1.157\times10^{-5}$Hz. Upper  frequency spectrum (red) is for detector in Mode A, i.e responding to 3-space dynamics and instrument noise, while lower spectrum (blue)  is for Mode B in which detector only responds to instrumental noise. We see that the signal in Mode A is very different from that Mode B operation, showing that the  S/N ratio for the detector is very high. The instrumental noise has  a mild `blue' noise spectrum, with  a small increase at higher frequencies, while the 3-space turbulence has a distinctive `pink' noise spectrum.}}
  \label{fig:FT}
\end{figure}

In Fig.\ref{fig:FT} the  Fourier Transfoms of the two data time series are shown, again revealing the very different characteristics of the data from the two operating modes. The instrumental noise has  a mild `blue' noise spectrum, with  a small increase at higher frequencies, while the 3-space turbulence has a distinctive `pink' noise spectrum, and ranging essentially from $0$ to $0.1$Hz.  The FT is defined by
\begin{equation}
\tilde{V}_s=\frac{1}{\surd{n}}\sum^n_{r=1}V_re^{2\pi i(r-1)(s-1)/n}
\label{eqn:FTdefn}\end{equation}
where $n=17280$ corresponds to a 5s timing interval over 24 hours.

\begin{figure}
\vspace{-7mm}\hspace{44mm} \includegraphics[scale=0.93]{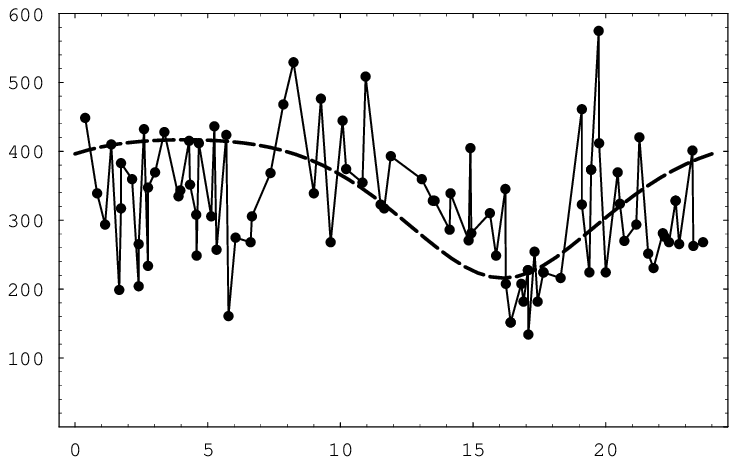}

\vspace{4mm} \hspace{36mm} \includegraphics[scale=0.25]{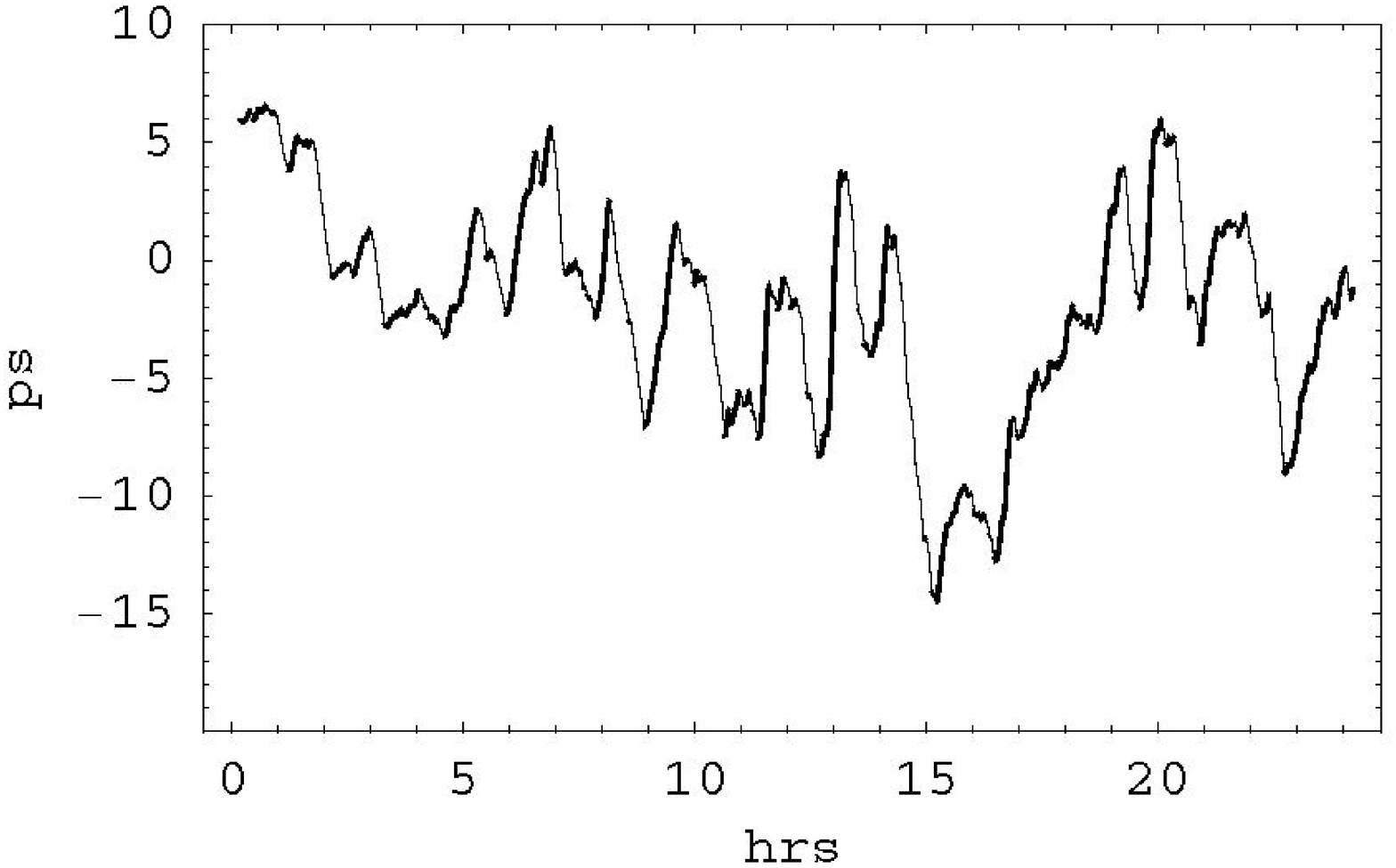}

\vspace{4mm} \hspace{42mm} \includegraphics[scale=0.35]{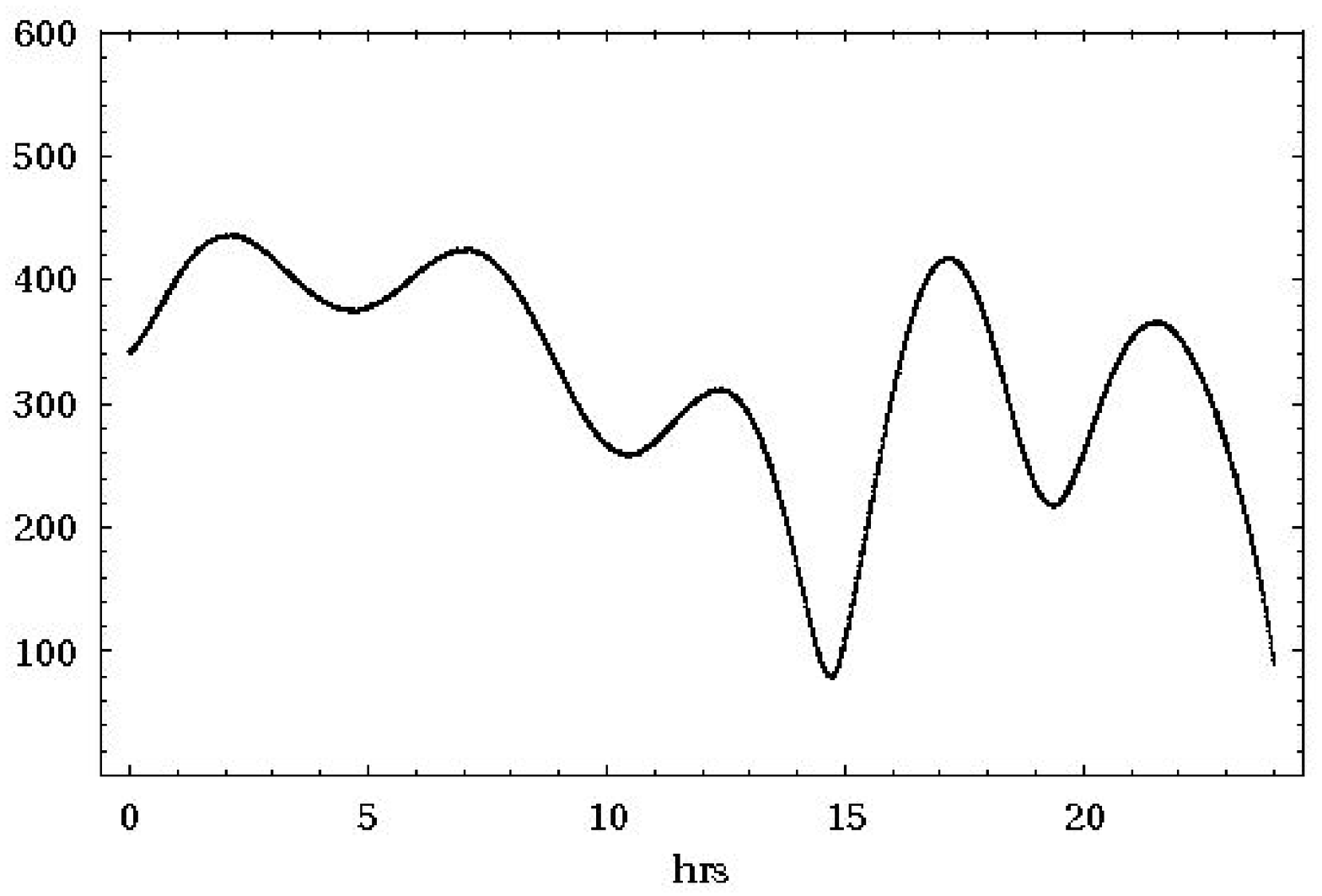}\caption{\small
{\it Top:} Absolute projected speeds $v_P$ in the Miller experiment  plotted against sidereal time in hours  for a composite day collected over a number of days in September 1925.   Maximum projected speed is 417\,km/s. The data shows considerable fluctuations.  The dashed curve shows the non-fluctuating  variation expected over one day as the Earth rotates, causing the projection onto  the plane of the interferometer of the  velocity of the average direction of the 3-space flow to change.    {\it Middle:}  Data from the Cahill experiment \cite{Coax}  for one sidereal day on approximately August 23, 2006.  We see similar variation with sidereal time, and also similar  wave structure.  This  data has been averaged over a running 1hr time interval to more closely match the time resolution of the Miller experiment. These fluctuations are  real wave phenomena of the 3-space.  {\it Bottom:} Data from the optical-fiber experiment herein with only low frequencies included to simulate the time averaging in the other two experiments. Comparison permits an approximate calibration for the optical fiber detector, as indicated by the speed in km/s. }
 \label{fig:SeptPlot}
\end{figure}

By removing all but the FT amplitudes 1-10, and then inverse Fourier Transforming we obtain the slow changes occurring over 24 hours. The resulting data has been presented in terms of possible values for the projected speed $v_P$ in (\ref{eqn:times}), and is shown in Fig.\ref{fig:SeptPlot}, after adjusting the unknown calibration constants to give a form resembling the Miller and coaxial cable experimental results, giving some indication of the calibration of the detector.  The experiment was run in an unoccupied office in which temperatures varied by some $10^0C$ over the 24 hour periods. In future   temperature control will be introduced.

\section{Conclusions}
As reviewed in \cite{Book,Review} gravitational waves, that is, fluctuations or turbulence in the dynamical 3-space, have been detected since  the 1887 Michelson-Morley experiment, although this all went unrealised until recently.
As the timing resolution improved over the century, from initially one hour to seconds now,  the characteristics of the turbulence of the dynamical 3-space have become more apparent, and that at smaller timing resolutions the turbulence is seen to be very large.  As shown herein this wave phenomenon is very easy to detect, and opens up a whole new window on the universe.  The detector reported here took measurements every 5s, but can  be run at  millisecond acquisition rates.  A 3D version of the detector with three orthogonal  optical-fiber interferometers will soon become operational. This will permit  the determination of the directional characteristics of the 3-space waves.  

That the average 3-space flow will affect  the gyroscope precessions in the GP-B satellite experiment through vorticity effects  was reported in \cite{GPB}.  The fluctuations are also predicted to be detectable in that experiment as noted in \cite{GPBwaves}.  However the much larger fluctuations detected in \cite{ Coax} and herein imply that these effects will be much large than reported in \cite{GPBwaves} where the time averaged waves from the DeWitte experiment  \cite{DeWitte} were used; essentially the gyro precessions will appear to have a large stochasticity.

Special thanks to Peter Morris, Thomas Goodey, Tim Eastman, Finn Stokes and  Dmitri Rabounski.

\end{document}